\documentstyle[11pt]{article}
\begin{document}
\centerline{\Large\bf Lorentz-Invariant Time-Energy Uncertainty Relation}
 \vskip 2mm
\centerline{\Large\bf for Relativistic Photon}
\vskip 2mm
\centerline{\bf S.N.Molotkov}
\centerline{\small\it Institute of Solid-State Physics, Russian Academy of Sciences}
\centerline{\small\it Chernogolovka, Moscow district, 142432 Russia}

\begin{abstract}
The time-energy uncertainty relation is discussed for a relativistic massless
particle. The Lorentz-invariant uncertainty relation is obtained between the
root-mean-square energy deviation and the scatter of registration time. The
interconnection between this uncertainty relation and its classical analogue is
established.
\end{abstract}
\vskip 2mm

The notion of time-energy uncertainty relation
\begin{equation}
\overline{(\Delta\varepsilon)^2}\cdot\overline{(\Delta t)^2}\ge
\frac{1}{4}
\end{equation}
in nonrelativistic quantum mechanics is not so well defined as
the other relations of this type, e.g., the coordinate-momentum
uncertainty relation [1--4]. This is primarily caused by the fact
that time is not a dynamical variable corresponding to a certain
Hermitian operator, but is a parameter. Because of the presence of
a lower bound in the spectrum of Hamiltonian, one generally cannot
introduce the Hermitian time operator [5]. The time-energy uncertainty
relation was discussed in many works for a great variety of situations.
For instance, in [6] the time-energy relation was derived for the
internal evolution of a quantum system, but it did not describe the
measurement process. A Hamiltonian allowing the instantaneous (in a
time as short as one likes), exact, and reproducible energy measurement
for a quantum system was written in [7]. True enough, no example of a
physical system is known so far to which this Hamiltonian could be applied.
In [7], the fact was used that external classical fields of a duration as
short as one likes and an intensity as high as one likes are allowed by
the formal apparatus of nonrelativistic quantum mechanics. This approach
has come to criticism in [8]. For the relativistic case, the restrictions
placed by special relativity on the measurability of quantum states were
first discussed in [9]. Further inquiry was undertaken in [10].
It turned out that, strictly speaking, only the classical fields
(potentials) can be treated classically in the Hamiltonian. The
time-dependent fields require quantum approach. Hence, the question
of exact and reproducible energy measurement in a time as short as one
likes was, in fact, merely reformulated in different terms.

Although time is not a dynamical variable, the measurement of event time
is a rather routine experimental situation [11]. Let the event time be
fixed experimentally; in this case, the registration time is a space of
results. The interrelation between the probability distribution on the
space of results (registration time) and the state of quantum system is
specified by a positive operator-valued measure. More precisely, to
every subset $\Delta_t\in(-\infty,\infty)$ of the space of results
there is a positive operator ${\cal M}(\Delta_t)$ such that
\begin{equation}
{\cal M}(\cup\Delta_{it})=\sum_i{\cal M}(\Delta_{it}),\quad
\Delta_{it}\cap\Delta_{jt}=\emptyset,
\end{equation}
The normalization condition is that the total probability of events
occurring over all space of results is unity:
\begin{equation}
{\cal M}(\Delta_{(-\infty,\infty)})=I,\quad
\Delta_{(-\infty,\infty)}\equiv(-\infty,\infty).
\end{equation}
In addition, the operator-valued measure in the identity resolution (2)
must satisfy the covariance condition. In the preparation of quantum state,
the shifts in time origin must lead to corresponding shifts in the
probability distribution; one has
\begin{equation}
\hat{U}_{t_0}{\cal M}(\Delta_t)\hat{U}^{-1}_{t_0}={\cal M}(\Delta_{t-t_0}),
\end{equation}
where $\hat{U}_t$ is the evolution (time shift) operator.
One can introduce the most symmetric time operator
\begin{equation}
\hat{t}=\int_{-\infty}^{\infty}t {\cal M}(t,dt).
\end{equation}
The mean registration time is given by the standard expression
\begin{equation}
\overline{t}=\int_{-\infty}^{\infty}t\mbox{Tr}\{{\cal M}(t,dt)\rho\},
\end{equation}
where $\rho$ is the density matrix of a quantum system subjected to measurement.

Accordingly, the root-mean-square deviation of registration time is defined as
\begin{equation}
\overline{(\Delta t)^2}=\int_{-\infty}^{\infty}(t-\overline{t})^2
\mbox{Tr}\{{\cal M}(t,dt)\rho\}.
\end{equation}

If $\hat{H}$ is the Hamiltonian of the system, then its spectral representation
is
\begin{equation}
\hat{H}=\int_{0}^{\infty}\varepsilon {\cal E}(\varepsilon,d\varepsilon),
\end{equation}
where ${\cal E}(\varepsilon,d\varepsilon)$ is the spectral family of
orthogonal projectors. Note that the operator-valued measures
${\cal M}(t,dt)$ in Eq. (2) are not orthogonal.
The mean energy and the root-mean-square deviation are defined for
the system in quantum state $\rho$ as
\begin{equation}
\overline{\varepsilon}=\int_{0}^{\infty}\varepsilon\mbox{Tr}\{
{\cal E}(\varepsilon,d\varepsilon)\rho\},
\end{equation}
\begin{equation}
\overline{(\Delta\varepsilon)^2}=\int_{0}^{\infty}
(\varepsilon-\overline{\varepsilon})^2\mbox{Tr}\{{\cal E}(\varepsilon,d\varepsilon)\rho\}.
\end{equation}

Next, one may raise the question as to the attainable lower bound of the time-energy
uncertainty relation, i.e., the question of what are the quantum states for which
the functional
\begin{equation}
\Omega=\min_{\{\rho\}}
\left\{
\overline{(\Delta\varepsilon)^2}\cdot\overline{(\Delta t)^2}
\right\}
\end{equation}
reaches its minimum.
Below, the time-energy relation in the sense given by Eqs.
(2--11) is considered for a one-dimensional massless relativistic
particle (photon). Although being model, this example, nevertheless,
encompasses all main features of the problem. Moreover, experiments with
photons, as a rule, are carried out in one-dimensional optical fiber systems.

Inasmuch as time is not an absolute category in the relativistic case,
the notion of time-energy uncertainty relation, at first glance, is defined
even worse than in the nonrelativistic case. However, the distinctive
feature of a photon is that its momentum and energy are linearly related
to each other. Moreover, since the mass shell of a massless field
coincides with the leading part of the light cone in momentum representation,
all events for the states propagating in one direction occur at the light cone
in the Minkowski space-time. As a result, the time-energy uncertainty relation
becomes Lorentz-invariant (independent of the inertial coordinate system where
the measurement is carried out). The lower bound in inequality (1) becomes
slightly higher than $1/4$.

Despite the fact that time is not an absolute category in the relativistic
case and that, in contrast to the nonrelativistic case, the notion of the
state (wave function) at a given instant of time (i.e., strictly speaking,
the Schr\"odinger representation) does not exist, the time-energy uncertainty
relation in the sense of Eqs. (2--11) is well defined.

The states of a free quantized field (more precisely, the generalized
eigenvectors) are generated by the action of field operators (generalized
functions with operator values)
\begin{equation}
\varphi^+(\hat{x})=\frac{1}{\sqrt{2\pi}}
\int d\hat{k}\delta(\hat{k}^2)\theta(k_0)\mbox{e}^{i\hat{k}\hat{x}}
a^+(\hat{k}),
\end{equation}
\begin{displaymath}
\hat{k}=(k,k_0),\quad\hat{x}=(x,t),
\quad \hat{k}=dkdk_0,\quad \hat{k}\hat{x}=kx-k_0t.
\end{displaymath}
on the vacuum vector [12]. The creation and annihilation
operators satisfy relations
\begin{equation}
[a^-(\hat{k}),a^+(\hat{k}')]=k_0\delta(k-k').
\end{equation}
The field physical states $|\psi\rangle\in{\cal H}$ belonging to the
Hilbert space of states are obtained by integrating the generalized
operator functions together with basic functions
$\psi(\hat{x})\in\Omega(\hat{x})$; the generalized eigenvectors
($\varphi^+(\hat{x})|0\rangle\in\Omega^*(\hat{x})$  are continuous
linear functionals on  $\Omega(\hat{x})$, where
$\Omega(\hat{x})\subset{\cal H}\subset\Omega^*(\hat{x})$
is the rigged Hilbert space [13] -- Gel'fand triple). One has
\begin{equation}
|\psi\rangle=\int d\hat{x}\psi(\hat{x})\varphi^+(\hat{x})|0\rangle=
\int d\hat{k}\psi(\hat{k})\delta(\hat{k}^2)\theta(k_0)a^+(\hat{k})|0\rangle=
\int_{-\infty}^{\infty}\frac{dk}{k_0}\psi(k,k_0=|k|)|\hat{k}\rangle,
\end{equation}
\begin{displaymath}
|\hat{k}\rangle=a^+(\hat{k})|0\rangle,\quad
\langle\hat{k}|\hat{k}'\rangle=k_0\delta(k-k'),\quad
\psi(\hat{k})=\int d\hat{x}\psi(\hat{x})\mbox{e}^{-i\hat{k}\hat{x}},
\end{displaymath}
where $dk/k_0$ the Lorentz-invariant integration volume.

The contribution to the physical state $|\psi\rangle$ comes from
the amplitude $\psi(k,k_0=|k|)$ at the mass shell (leading part
of the light cone in momentum representation).

We consider the states propagating in one direction. For the states
propagating in both directions, the notion of event time has no sense.
For the states propagating in one direction (for definiteness, $k>0$),
energy and momentum are one and the same, because of the linear relationship
between them, $k_0=|k|=k$. For these states, only the vectors with $k > 0$
make contribution to Eq. (14), and the amplitude $\psi(k,k)$ is nonzero at
$k > 0$.

The energy (momentum) measurement is given by the identity resolution in the
subspace of one-particle states:
\begin{equation}
I=\int_{-\infty}^{\infty}\frac{dk}{k_0}|\hat{k}\rangle\langle\hat{k}|=
\int_{-\infty}^{\infty}{\cal M}(k,dk),\quad
{\cal M}(k,dk)=|\hat{k}\rangle\langle\hat{k}|\frac{dk}{k_0},
\quad I_+=\int_{0}^{\infty}{\cal M}(k,dk).
\end{equation}
In actuality, it will suffice to restrict oneself to the subspace of states
projected onto the vectors $|\hat{k}\rangle$ with
$k>0$, $I_+$ is unity in this subspace.
The probability of measuring energy (momentum) in the interval $(k,k+dk)$
is given by
\begin{equation}
\mbox{\bf Pr}\{dk\}=\mbox{Tr}\{{\cal M}(k,dk)|\psi\rangle\langle\psi|\}=
|\psi(k,k)|^2\frac{dk}{k}=|f(k)|^2dk,\quad
f(k)=\frac{\psi(k,k)}{ \sqrt{k} }.
\end{equation}
The mean energy (momentum) in the state $|\psi\rangle$ is
\begin{equation}
\overline{k}=\int_{0}^{\infty}k\mbox{\bf Pr}\{dk\}=
\int_{0}^{\infty}k|f(k)|^2dk,
\end{equation}
and the root-mean-square deviation is
\begin{equation}
\overline{(\Delta k)^2}=\int_{0}^{\infty}(k-\overline{k})^2
\mbox{\bf Pr}\{dk\}=\overline{k^2}-(\overline{k})^2,\quad
\overline{k^2}=\int_{0}^{\infty}k^2\mbox{\bf Pr}\{dk\}.
\end{equation}
Let us now consider the measurement of particle position; for the
states propagating in one direction ($k>0$), this position can be
represented by the expansion of unity:
\begin{equation}
I_+=\int_{-\infty}^{\infty}\frac{dx}{2\pi}
\left(\int_{0}^{\infty}\frac{dk}{\sqrt{k}}\mbox{e}^{-i\hat{k}\hat{x}}
|\hat{k}\rangle\right)
\left(\int_{0}^{\infty}\frac{dk'}{\sqrt{k'}}\langle\hat{k'}|
\mbox{e}^{i\hat{k}'\hat{x}}\right)=
\end{equation}
\begin{displaymath}
\int_{-\infty}^{\infty}\frac{d\tau}{2\pi}
\left(\int_{0}^{\infty}\frac{dk}{\sqrt{k}}\mbox{e}^{-ik\tau}
|\hat{k}\rangle\right)
\left(\int_{0}^{\infty}\frac{dk'}{\sqrt{k'}}\langle\hat{k'}|
\mbox{e}^{ik'\tau}\right)=
\int_{-\infty}^{\infty}{\cal M}(\tau,d\tau),
\end{displaymath}
where $\tau=x-t$. The measurement of the coordinate $x$ is,
in fact, the measurement of response time $t$. More precisely,
the space of results is formed not by $x$ and $t$ separately, but
by their difference $\tau$. The expansion of unity in Eq. (19) formally
describes a device; it can be interpreted as follows. If the
space of results is formed by $x$, then the measurement should be
understood as an $x$-distributed device generating random result
at point $(x,x+dx)$ and time $t$. If $x$ is fixed, then the measurement
describes an $x$-localized instrument operating
in a trigger mode and generating result at a random instant of time.
The fact that the operator-valued measure ${\cal M}(\tau,d\tau)$
in Eq. (19) depends only on the difference $\tau=x-t$ means that,
if the result can be obtained at the point $x$ at the time instant $t$ with
a certain probability, then the same result can be obtained at a different
point $x$' with the same probability, but at the instant of time
$t'=x'-x+t$.

Accordingly, the probability to obtain the result at time interval
$(\tau,\tau+\tau)$ is, by definition,
\begin{equation}
\mbox{\bf Pr}\{d\tau\}=\mbox{Tr}\{{\cal M}(\tau,d\tau)|\psi\rangle\langle\psi|\}
=|f(\tau)|^2d\tau,
\end{equation}
\begin{equation}
f(\tau)=\int_{0}^{\infty}\frac{dk}{\sqrt{k}}\psi(k,k)\mbox{e}^{-ik\tau}.
\end{equation}
It is notable that $f(\tau)$ coincides with the Landau-Peierls wave
function in coordinate representation [14]. Contrary to $\overline{k}$, the
mean value of $\overline{\tau}$ can be chosen to be zero upon the
appropriate choice of time origin. The root-mean-square deviation of
registration time is
\begin{equation}
\overline{(\Delta\tau)^2}=\int_{-\infty}^{\infty}d\tau
\mbox{\bf Pr}\{ d\tau \}=\int_{-\infty}^{\infty}\tau^2|f(\tau)|^2d\tau=
\end{equation}
\begin{displaymath}
\int_{0}^{\infty}\int_{0}^{\infty}dkdk'f(k)f^*(k')\int_{-\infty}^{\infty}\tau^2
\mbox{e}^{i(k-k')\tau }d\tau=
\int_{0}^{\infty}\int_{0}^{\infty}dkdk'f(k)f^*(k')
\frac{\partial^2}{\partial k\partial k'}\delta(k-k')=
\end{displaymath}
\begin{displaymath}
\int_{0}^{\infty}\Big|\frac{df(k)}{dk}\Big|^2dk.
\end{displaymath}
The further goal consists of finding the state $|\psi\rangle$,
for which the functional
\begin{equation}
\Omega(f)=\left(
\int_{0}^{\infty}\Big|\frac{df(k)}{dk}\Big|^2dk
\right)
\left(
\int_{0}^{\infty}(k^2-\overline{k}^2)|f(k)|^2dk
\right),
\end{equation}
is minimum under the additional normalization condition
\begin{equation}
\langle\psi|\psi\rangle=\int_{0}^{\infty}\frac{dk}{k}|\psi(k,k)|^2=
\int_{0}^{\infty}|f(k)|^2dk=1.
\end{equation}
It turns out that the problem
\begin{equation}
\frac{\delta\Omega(f)}{\delta f}=0
\end{equation}
of minimizing functional (23) was solved for
classical signals in the elegant though little-known work [15] as
early as 1934 (see also [16,17]). It was shown that, for the
time-frequency uncertainty relation defined as in Eq. (23), the
functional reaches its minimum on even time functions $f(\tau)$
(accordingly, $df(k)/dk|_{k=0}=0$). The corresponding variational
problem reduces to the second-order differential equation for
$f(k)$ of the form
\begin{equation}
\frac{d^2f(x)}{dx^2}+\left(\nu+\frac{1}{2}-\frac{x^2}{4}\right)f(x)=0,
\quad
x=\left( \frac{4a}{b-c^2}\right)^{1/4}(k-c),\quad
\nu+\frac{1}{2}=\sqrt{a(b-c^2)},
\end{equation}
where
\begin{equation}
a=\int_{0}^{\infty}\left(\frac{df(k)}{dk}^2\right)^2dk,
\quad
b=\int_{0}^{\infty}k^2f(k)^2dk,
\quad
c=\int_{0}^{\infty}kf(k)^2dk,
\quad
\int_{0}^{\infty}f^2(k)dk=1,
\end{equation}
Here, the integrals $a,b,c$ are taken along the extremum.
Taking into account that $df(k)/dk|_{k=0}=0$ and that
$b=3c^2/2$ for the extremum, the solution is given by the
parabolic cylinder function (Weber function) $D_{\nu}(x)$ [18].
The value of $\nu$ is determined from the condition
$D'_{\nu}(x)=dD_{\nu}(x)dx=0$. Taking into account that
\begin{equation}
D_{\nu}(x)=\frac{\mbox{e}^{-x^2/4}}{\Gamma(-\nu)}
\int_{0}^{\infty}\mbox{e}^{-x\xi-\xi^2/2}\xi^{-\nu-1}d\xi,
\end{equation}
this is equivalent to the solution of the transcendental equation
\begin{equation}
D'_{\mu-\frac{1}{2}}(-2\sqrt{\mu})=0 ,
\quad
\int_{0}^{\infty}\mbox{e}^{2\sqrt{\mu}\xi-\xi^2/2}\xi^{-\mu-1/2}
(\xi-\sqrt{\mu})d\xi=0.
\end{equation}
where $\mu=\nu+1/2$. The numerical value is given in [17] :
$\mu^2=0.2951..$. The functional in its extremum equals
\begin{equation}
\Omega_{min}(f)=\frac{a\cdot b}{3}=\mu^2=0.2951..
\end{equation}
Let us now show that these time-energy uncertainty
relations are Lorentz-invariant, i.e., remain unchanged upon measuring
quantum state in any inertial frame of reference. The measurements in
the observer's frame of reference themselves are formulated as in Eqs. (15)
and (19); in doing so, by all quantities in Eqs. (15) and (19) should be
meant their values in the observer's frame of reference, while the state
obtained by the action of the respective unitary operator of the Poincare
group representation should be taken as a quantum state ``seen''
by the observer in the moving coordinate system. The general coordinate
transformation in the Poincare group consists of the translation in the
Minkowski space-time and the Lorentzian rotation; one has
\begin{equation}
\hat{x}'=\hat{P}(\hat{a})\hat{L}\hat{x}=\hat{L}\hat{x}+\hat{a},
\end{equation}
where $\hat{P}(\hat{a})$ is the operator of translation by
$\hat{a}=(a,a_0)$ and $\hat{L}$ is the operator of Lorentzian
rotation describing the transition to a different inertial system.
These transformations induce operator transformations
\begin{equation}
\hat{\mbox{\bf U}}(\hat{L},\hat{a})a^+(\hat{k})
\hat{\mbox{\bf U}}^{-1}(\hat{L},\hat{a})=
\mbox{e}^{i\hat{L}\hat{k}\cdot\hat{a}}a^+(\hat{L}\hat{k}),
\end{equation}
where $\hat{\mbox{\bf U}}(\hat{L},\hat{a})$ is the unitary operator acting
in ${\cal H}$.

The transformed state effectively seen by the observer is
\begin{equation}
|\psi(\hat{L},\hat{a})\rangle=\hat{\mbox{\bf U}}(\hat{L},\hat{a})|\psi\rangle=
\end{equation}
\begin{displaymath}
\int d\hat{x}\psi(\hat{x})\hat{\mbox{\bf U}}(\hat{L},\hat{a})
\varphi^+(\hat{x})\hat{\mbox{\bf U}}(\hat{L},\hat{a})^{-1}
\hat{\mbox{\bf U}}(\hat{L},\hat{a})|0\rangle=
\end{displaymath}
\begin{displaymath}
\int d\hat{x}\psi(\hat{x})\varphi(\hat{L}\hat{x}+\hat{a})|0\rangle=
\int d\hat{x}\psi(\hat{L}^{-1}(\hat{x}-\hat{a}))\varphi^+(\hat{x})|0\rangle=
\end{displaymath}
\begin{displaymath}
\int d\hat{k}
\psi(\hat{k})\mbox{e}^{i\hat{k}\hat{k}\cdot\hat{a}}\delta(\hat{k}^2)
\theta(k_0)a^+(\hat{L}\hat{k})|0\rangle=
\int d\hat{k}\psi(\hat{L}^{-1}\hat{k})
\mbox{e}^{i\hat{k}\hat{a}}|\hat{k}\rangle=
\end{displaymath}
\begin{displaymath}
\int_{0}^{\infty}\frac{dk}{k}
\psi(\hat{L}^{-1}\hat{k})\mbox{e}^{i\hat{k}\hat{a}}
|\hat{k}\rangle=
\int_{0}^{\infty}\frac{dk}{k}
\psi\left( \frac{k-\beta k_0}{\sqrt{1-\beta^2}}
,\frac{k_0-\beta k}{\sqrt{1-\beta^2}}
\right) \mbox{e}^{i(ka-k_0a_0)}|\hat{k}\rangle,
\end{displaymath}
where $dk/k_0$ is the Lorentz-invariant volume of integration. In Eq. (33),
the invariance of the vacuum vector,
$\hat{\mbox{\bf U}}(\hat{L},\hat{a})|0\rangle=|0\rangle$, is also taken into
account. Recall that only those states are considered which propagate in
one direction along the $x$ axis. The final state seen by the observer is
written as
\begin{equation}
|\psi(\hat{L},\hat{a})\rangle=\int_{0}^{\infty} \frac{dk}{k}\psi\left(
k\sqrt{\frac{1-\beta}{1+\beta}},
k\sqrt{\frac{1-\beta}{1+\beta}}
\right)
|\hat{k}\rangle.
\end{equation}
The mean energy (momentum) measured by the observer is defined as
(the quantities in the moving coordinate system are labeled $m$)
\begin{equation}
\overline{k_m}=\int_{0}^{\infty}k\mbox{\bf Pr}\{dk\}=\int_{0}^{\infty}
k\mbox{Tr}\{{\cal M}(k,dk)
|\psi(\hat{L},\hat{a})\rangle\langle\psi(\hat{L},\hat{a}|\}
=
\end{equation}
\begin{displaymath}
\int_{0}^{\infty}\frac{dk}{k}k
\Big|\psi\left(k\sqrt{\frac{1-\beta}{1+\beta}},
k\sqrt{\frac{1-\beta}{1+\beta}}\right)\Big|^2=
\overline{k}\sqrt{\frac{1+\beta}{1-\beta}}.
\end{displaymath}
At small $\beta\ll1$, the mean momentum (energy) in the moving system
is related to its value in the fixed coordinate system by the expression
\begin{equation}
\overline{k_m}=\overline{k}(1+\beta)=k(1+\frac{v}{c}),
\end{equation}
which, in fact, is a formulation of the Doppler effect.
The respective energy (momentum) root-mean
square deviation in the moving coordinate system is
\begin{equation}
\overline{(\Delta k)^2_m}=\int_{0}^{\infty}(k-\overline{k_m})^2
\mbox{\bf Pr}\{dk\}=
\int_{0}^{\infty} (k-\overline{k_m})^2
\mbox{Tr}\{{\cal M}(k,dk)
|\psi(\hat{L},\hat{a})\rangle\langle\psi(\hat{L},\hat{a}|\}=
\end{equation}
\begin{displaymath}
\int_{0}^{\infty}\frac{dk}{k} (k-\overline{k_m})^2
\Big| \psi\left( k\sqrt{ \frac{1-\beta}{1+\beta} },
k\sqrt{ \frac{1-\beta}{1+\beta} } \right)\Big|^2=
\left( \frac{1+\beta}{1-\beta} \right)
\overline{(\Delta k)^2}.
\end{displaymath}
The root-mean square deviation of registration time in the moving system
is defined as (for the sake of convenience, the coordinate systems in this
formula have the common origin, i.e., $\hat{a}=0$, and only the Lorentzian
rotation $\hat{L}$ is retained)
\begin{equation}
\overline{(\Delta \tau)_m^2}=\int_{-\infty}^{\infty}\tau^2\mbox{\bf Pr}\{d\tau\}=
\int_{-\infty}^{\infty}\tau^2\mbox{Tr}\{{\cal M}(\tau,d\tau)|\psi(\hat{L},\hat{0})\rangle
\langle\psi(\hat{L},\hat{0}|\}=
\end{equation}
\begin{displaymath}
\int_{-\infty}^{\infty}\tau^2
\Big|\int_{0}^{\infty}\frac{dk}{\sqrt{k}}
\mbox{e}^{-ik\tau}\psi\left(k
\sqrt{\frac{1-\beta}{1+\beta}},\sqrt{\frac{1-\beta}{1+\beta}}
\right)\Big|^2\frac{d\tau}{2\pi}=
\end{displaymath}
\begin{displaymath}
\int_{-\infty}^{\infty}\tau^2
\Big|\int_{0}^{\infty}\frac{dk}{\sqrt{k}}
\mbox{e}^{-ik\left( \tau\sqrt{ \frac{1+\beta}{1-\beta} } \right)}
\psi(k,k) \Big|^2
d\left( \frac{\tau}{2\pi}\sqrt{\frac{1+\beta}{1-\beta} } \right)=
\overline{(\Delta\tau)^2}\left(\frac{1-\beta}{1+\beta}\right).
\end{displaymath}
It follows from Eqs. (37) and (38) that the resulting time-energy
uncertainty relation in the observer's moving coordinate system
is related to the uncertainty relation in the initial system as
\begin{equation}
\overline{(\Delta k)^2_m}\cdot\overline{(\Delta\tau)^2_m}=
\overline{(\Delta k)^2}\cdot\overline{(\Delta\tau)^2}=0.2951..>\frac{1}{4}.
\end{equation}
i.e., it is Lorentz-invariant.

The fact that the time-energy uncertainty relation is Lorentz-invariant
is due, in fact, to the covariance of the energy (momentum) and event-time
measurements. Indeed, the operator-valued measure in Eq. (19) is covariant
about the Poincare group transformations:
\begin{equation}
\hat{\mbox{\bf U}}(\hat{L},\hat{a}){\cal M}(\tau,d\tau)
\hat{\mbox{\bf U}}(\hat{L},\hat{a})^{-1}=
{\cal M}\left(\sqrt{\frac{1-\beta}{1+\beta}}\tau-(a-a_0),
d \left(\sqrt{\frac{1-\beta}{1+\beta}}\tau\right)\right).
\end{equation}
The momentum measurement and the orthogonal operator measure in Eq. (15)
also satisfy the covariance condition
\begin{equation}
\hat{\mbox{\bf U}}(\hat{L},\hat{a}){\cal M}(k,dk)
\hat{\mbox{\bf U}}(\hat{L},\hat{a})^{-1}=
{\cal M}\left(\sqrt{\frac{1+\beta}{1-\beta}}k,d
\left(\sqrt{\frac{1+\beta}{1-\beta}}k\right)\right).
\end{equation}
If the measurement occurs in the same inertial coordinate system,
$\hat{L}=\hat{1}$, then the covariance condition (41) is analogous
to the nonrelativistic case (4) with the only difference that the
covariance is understood in the sense of translations in the
Minkowski space-time (in our case, shifts along the light cone
branch).

I am grateful to S.S.Nazin for discussions and critical remarks.

This work was supported by the projects ``Physical Foundations of
Quantum Computer'' and ``Electronic States''.


\begin{thebibliography}{99}
\bibitem{1}W.Heisenberg, Z. Phys., {\bf 60}, 56 (1927).
\bibitem{2}N.Bohr, Selected Scientific Works (Nauka, Moscow, 1971 ), Vol. 2.
\bibitem{3}N.S.Krylov and V.A.Fok, Zh. Eksp. Teor. Fiz. {\bf 17}, 93 ( 1947).
\bibitem{4}E.P.Wigner, {\it On the time-energy uncertainty relation}, in
{\it Aspects of Quantum Theory}, ed.A.Salam, E.P.Wigner, Cambridge University
Press, Mass., 237 (1972).
\bibitem{5}A.S.Holevo, {\it Probabilistic and Statistical Aspects of Quantum Theory},
North-Holland, Amsterdam, 1980.
\bibitem{6}L.I.Mandel'shtam and I.E.Tamm, Izv. Akad. Nauk  SSSR, Ser. Fiz. {\bf 9}, 122 (1945).
\bibitem{7}Y.Aharonov, D.Bohm, Phys. Rev., {\bf 122}, 1649 (1961); Y.Aharonov, J.L.Safko,
Ann. Phys., {\bf 91}, 279 (1975).
\bibitem{8}V.A.Fok, Zh. Eksp. Teor Fiz. {\bf 42}, 1135 (1962)
[Sov.  Phys. JETP {\bf 15}, 784 ( 1962)].
\bibitem{9}L.D.Landau and R.Peierls, Z. Phys. {\bf 69}, 56 (1931);
L.D.Landau, {\it A Collection of Scientific Works} (Nauka, Moscow,
1969), Vol.1, p.56.
\bibitem{10}N.Bohr and L.Rosenfeld, Math.-Fys. Medd. {\bf 12}, 3  (1933);
N.Bohr, {\it A Collection of Scientific Works} (Nauka,  Moscow,
1971).
\bibitem{11}P.Busch, {\it The Time Energy Uncertainty Relation},
quant-ph/0105049; P.Busch, M.Grabowski, P.J.Lahti, {\it Operational Quantum
Physics}, Springer Lecture Notes in Physics, v.{\bf 31}, 1995.
\bibitem{12}N.N.Bogolyubov, A.A.Logunov, A.I.Oksak, and I.T.Todorov,
{\it General Principles of Quantum Field Theory} (Nauka, Moscow, 1987).
\bibitem{13}I.M.Gel'fand and N.Ya.Vilenkin, {\it Generalized Functions}
(Fizmatgiz, Moscow, 1961; Academic, New York, 1964), Vol.4.
\bibitem{14}L.D.Landau and R.Peierls, Z. Phys. {\bf 62}, 188 ( 1930);
L.D.Landau, {\it A Collection of Scientific Works} (Nauka, Moscow,
1969), Vol.1, p.32.
\bibitem{15}A.G.Mayer and E.A.Leontovich, Dokl. Akad. Nauk SSSR {\bf 4}, 353 (1934).
\bibitem{16}A.A.Kharkevich, {\it Spectra and Analysis} (Fizmatgiz, Moscow, 1962).
\bibitem{17}V.V.Dodonov and V.I.Man'ko, in {\it Invariants and Evolution of
Nonstationary Quantum Systems}, Ed. by M.A.Markov (Nauka, Moscow,
1987), Tr. Fiz. Inst. Akad. Nauk SSSR, Vol.183.
\bibitem{18}{\it Heigher Transcendental Functions (Bateman Manuscript Project)},
Ed. by A.Erdelyi (McGraw-Hill, New York, 1953; Nauka, Moscow, 1966).
\end{thebibliography}
\end{document}